\documentstyle[epsfig]{mn}

\newif\ifAMStwofonts

\ifoldfss
  \ifCUPmtlplainloaded \else
    \NewTextAlphabet{textbfit} {cmbxti10} {}
    \NewTextAlphabet{textbfss} {cmssbx10} {}
    \NewMathAlphabet{mathbfit} {cmbxti10} {} 
    \NewMathAlphabet{mathbfss} {cmssbx10} {} 
  \fi
  \ifAMStwofonts
    \ifCUPmtlplainloaded \else
      \NewSymbolFont{upmath} {eurm10}
      \NewSymbolFont{AMSa} {msam10}
      \NewMathSymbol{\upi}     {0}{upmath}{19}
      \NewMathSymbol{\umu}     {0}{upmath}{16}
      \NewMathSymbol{\upartial}{0}{upmath}{40}
      \NewMathSymbol{\leqslant}{3}{AMSa}{36}
      \NewMathSymbol{\geqslant}{3}{AMSa}{3E}

    \fi
  \fi
\fi 

\ifnfssone
  \newmathalphabet{\mathit}
  \addtoversion{normal}{\mathit}{cmr}{m}{it}
  \addtoversion{bold}{\mathit}{cmr}{bx}{it}
  \newmathalphabet{\mathbfit} 
  \addtoversion{normal}{\mathbfit}{cmr}{bx}{it}
  \addtoversion{bold}{\mathbfit}{cmr}{bx}{it}
  \newmathalphabet{\mathbfss} 
  \addtoversion{normal}{\mathbfss}{cmss}{bx}{n}
  \addtoversion{bold}{\mathbfss}{cmss}{bx}{n}
  \ifAMStwofonts
    \ifCUPmtlplainloaded \else
      \UseAMStwoboldmath
      \makeatletter
      \new@mathgroup\upmath@group
      \define@mathgroup\mv@normal\upmath@group{eur}{m}{n}
      \define@mathgroup\mv@bold\upmath@group{eur}{b}{n}
      \edef\UPM{\hexnumber\upmath@group}
      \new@mathgroup\amsa@group
      \define@mathgroup\mv@normal\amsa@group{msa}{m}{n}
      \define@mathgroup\mv@bold\amsa@group{msa}{m}{n}
      \edef\AMSa{\hexnumber\amsa@group}
      \makeatother
      \mathchardef\upi="0\UPM19
      \mathchardef\umu="0\UPM16
      \mathchardef\upartial="0\UPM40
      \mathchardef\leqslant="3\AMSa36
      \mathchardef\geqslant="3\AMSa3E
    \fi
  \fi
\fi 

\ifnfsstwo
  \DeclareMathAlphabet{\mathbfit}{OT1}{cmr}{bx}{it}
  \SetMathAlphabet\mathbfit{bold}{OT1}{cmr}{bx}{it}
  \DeclareMathAlphabet{\mathbfss}{OT1}{cmss}{bx}{n}
  \SetMathAlphabet\mathbfss{bold}{OT1}{cmss}{bx}{n}
  \ifAMStwofonts
    \ifCUPmtlplainloaded \else
      \DeclareSymbolFont{UPM}{U}{eur}{m}{n}
      \SetSymbolFont{UPM}{bold}{U}{eur}{b}{n}
      \DeclareSymbolFont{AMSa}{U}{msa}{m}{n}
      \DeclareMathSymbol{\upi}{0}{UPM}{"19}
      \DeclareMathSymbol{\umu}{0}{UPM}{"16}
      \DeclareMathSymbol{\upartial}{0}{UPM}{"40}
      \DeclareMathSymbol{\leqslant}{3}{AMSa}{"36}
      \DeclareMathSymbol{\geqslant}{3}{AMSa}{"3E}
    \fi
  \fi
\fi 

\ifCUPmtlplainloaded \else
  \ifAMStwofonts \else 
    \def\upi{\pi}
    \def\umu{\mu}
    \def\upartial{\partial}
  \fi
\fi

\title[Radial velocities in PG~1605$+$072]
{Radial velocity variations of the pulsating subdwarf B star
PG~1605$+$072\thanks{Based on observations made with the Anglo-Australian
Telescope, Coonabarabran, NSW, Australia and the William Herschel
Telescope operated on the island of La Palma by the Isaac Newton Group in the
Spanish Observatorio del Roque de los Muchachos of the Instituto de
Astrofisica de Canarias.  }}
\author[V. M. Woolf et al.]
       {Vincent M. Woolf,$^1$\thanks{email: vmw@star.arm.ac.uk}
 C. Simon Jeffery,$^1$ and Donald L.  Pollacco,$^{2,3}$\\
        $^1$Armagh Observatory, College Hill, Armagh BT61 9DG,
         Northern Ireland \\
        $^2$Isaac Newton Group, La Palma, Apartado de correos 321,
          E-38700 Santa Cruz de la Palma, Tenerife, Spain \\
        $^3$Queens University Belfast, Belfast BT7 1NN, Northern Ireland}
\date{}

\pubyear{2001}

\begin{document}

\maketitle

\label{firstpage}

\begin{abstract}
We present an analysis of high-speed spectroscopy of the pulsating subdwarf B
star PG~1605$+$072.  Periodic radial motions are detected at frequencies
similar to those reported for photometric variations in the star, with
amplitudes of up to 6 ${\rm km\, s^{-1}}$.  Differences between relative
strengths for given frequency peaks for our velocity data and previously
measured photometry are probably a result of shifting of power between
modes over time. Small differences in the detected frequencies may also
indicate mode-shifting.  We report the detection of line-shape variations
using the moments of the cross correlation function profiles.  It may be
possible to use the moments to identify the star's pulsation modes.
\end{abstract}


\section{Introduction}

It has been several years since the announcement of the discovery that
some subdwarf B stars pulsate \cite{k97}.  Photometric variations in sdBV stars,
or ``EC~14026 stars'' after the prototype, are measured in hundredths of
a magnitude and generally have periods of 100--200 s. They
appear to be due to low-order stellar pulsations.  When the pulsations are
better understood, it will be possible to use asteroseismology to
obtain more information about the stars, including their size, mass, and
structure, as has been done with some other classes of pulsating stars.
This will probably require multi-site photometric and spectroscopic
observations.

PG~1605$+072$ has the largest photometric pulsation amplitudes and the
richest frequency spectrum of all studied sdBV's.  Its light curve shows more
than 50 frequencies, though it is dominated by 5 frequencies between 1.89 and
2.74 mHz (periods between 529 and 365 s) \cite{ko98,k99}.  It has an effective
temperature of $32\,300 \pm 300$~K (Heber, Reid, \& Werner 1999).
It has a lower gravity ($\log g = 5.25$) and longer pulsation periods than 
any other sdBV studied, which Kilkenny et~al. \shortcite{k99} say indicates
that it has evolved away from the core helium burning horizontal branch
where the shorter-period sdBV's are found.  It is expected to be evolving
more rapidly than other known sdBV's.

Kawaler \shortcite{kw99} and Kilkenny et~al. \shortcite{k99} report comparisons
of the large-amplitude pulsations of PG~1605+072 to those of a model with
similar physical parameters.  They suggest that what are observed are most
likely low-order nonradial pulsations in ``trapped modes,''  though it
was not possible to make individual mode identifications.

An earlier study, O'Toole et al. \shortcite{o00}, used time-resolved
spectroscopy of PG~1605$+$072 to look for evidence of pulsation in the radial
velocities. Their observations provided 38.65 hours of data over a 10 day
period.  Spectra within a data series were generally separated by 61 to 75
seconds.  They detected velocity variations, with
the three largest amplitudes having frequencies similar to those found
in the photometric data by Kilkenny et al. \shortcite{k99}.  The combination
of the fairly slow repetition of observations, noise in the velocity spectrum
caused by the use of small diameter ($< 2$-m) telescopes,
and a broad observational alias
signature in comparison with the spacing of the frequencies observed in the
photometry meant that only the strongest  pulsations could be detected with
any certainty. 

We report the results of two-site spectroscopic observations of PG~1605$+$072.
We take advantage of the higher signal to noise spectra and faster
readout repetition possible with the 4-m class Anglo-Australian (AAT) and
William Herschel (WHT) telescopes to provide radial velocities for a study of
the star's pulsations.  Previous work has shown that high speed
spectroscopy at the WHT can be used to detect even small amplitude radial
motions, ${\rm \sim 2\, km\, s^{-1}}$, in sdBVs \cite{jp00}.

\section{Observations and reduction}

We obtained spectra of PG~1605$+$072 at a high time resolution (13 to 22
second repetition)
using the 3.9-m AAT and the 4.2-m WHT on the nights of 11 and 12
May 2000.  Using both telescopes allowed us to obtain a more continuous
data set than possible with a single site.  We obtained 16.3 hours of data
over a 32.1 hour period.  Spectra from both sites covered
a 400~\AA\ range chosen to include the H$\delta$ and H$\gamma$ Balmer lines.

The data were reduced using scripted standard {\sc iraf} routines to to make
the bias, flat-field, and sky corrections, extract the one dimensional
spectra, and to apply the wavelength calibration from CuAr arc spectra.

Exposure timings at both sites are calibrated against standard clocks. The 1 or
2 second possible time calibration difference between sites has no detectable
effect on the results we report.

\subsection{WHT observations}

We used the blue arm of the ISIS spectrograph with the
R1200B grating.  The CCD was read out
in drift mode \cite{r97}. This allowed us to obtain spectra
every 13 seconds. Observations were made over a period of 43 minutes
starting at  ${\rm HJD = 2451676.40761}$ and a period of 8.349 hours starting
at ${\rm HJD = 2451677.39750}$, with short breaks for CuAr arc calibration
spectra. Spectral resolution was $\lambda / \Delta \lambda = 5000$ and the
spectral range was ${\rm 4050 < \lambda < 4450\AA }$.  The exposure timings
come from the CCD controller's clock, which is synchronized daily with the
observatory's GPS-calibrated time service.  The clock normally drifts about 1
second per night.

\subsection{AAT observations}

We used the Royal Greenwich Observatory Spectrograph with CCD readout in
time-series mode \cite{sj97}.
We obtained a spectrum every 22.5 seconds over a period of
7.272 hours starting at ${\rm HJD = 2451676.97869}$, with short breaks every
30 minutes so CuAr arc spectra could be measured for wavelength calibration.
Using the 82 cm camera and the R1200B grating gave a spectral resolution of 
$\lambda / \Delta \lambda = 5700$ with a spectral range of
${\rm 4010 < \lambda < 4410\AA }$.  The exposure timings are calibrated against
the Observatory's CAMAC clock and should be accurate to within 20 ms.

\section{Analysis and Results}
The determination of radial velocities from the spectra and the search for
periods in the velocity variations followed the method outlined by Jeffery
\& Pollacco \shortcite{jp00} using {\sc idl} programs.

\begin{figure}
\epsfig{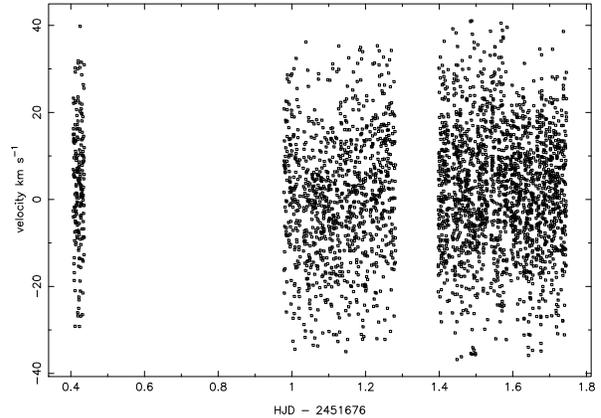}
\caption{Radial velocities measured for PG~1605$+$072.}
\label{figa}
\end{figure}

\begin{figure}
\epsfig{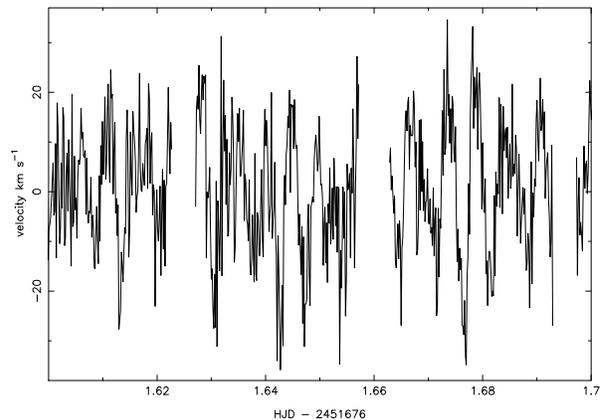}
\caption{Expanded section of the radial velocity curve.}
\label{figb}
\end{figure}

Wavelength calibrated spectra were used to determine radial velocity changes
of PG~1605$+$072.  As the spectral wavelength coverage from the two telescopes
was slightly different, data from each site were treated separately to simplify
the analysis.  For each site's data set the mean spectrum was used as the
cross-correlation template. 
The velocity for each spectrum was determined by fitting a Gaussian
to the the peak of the cross-correlation function.
The templates from the two data sets were cross-correlated to allow a correction
to be made to put the data sets on the same zero-scale.
The velocity data is shown in Fig.~\ref{figa}.
The velocity variations can be clearly seen in an expanded portion of the
velocity curve shown in Fig.~\ref{figb} (as can the gaps in the data due to
the need to take wavelength calibration arcs).

\begin{figure}
\epsfig{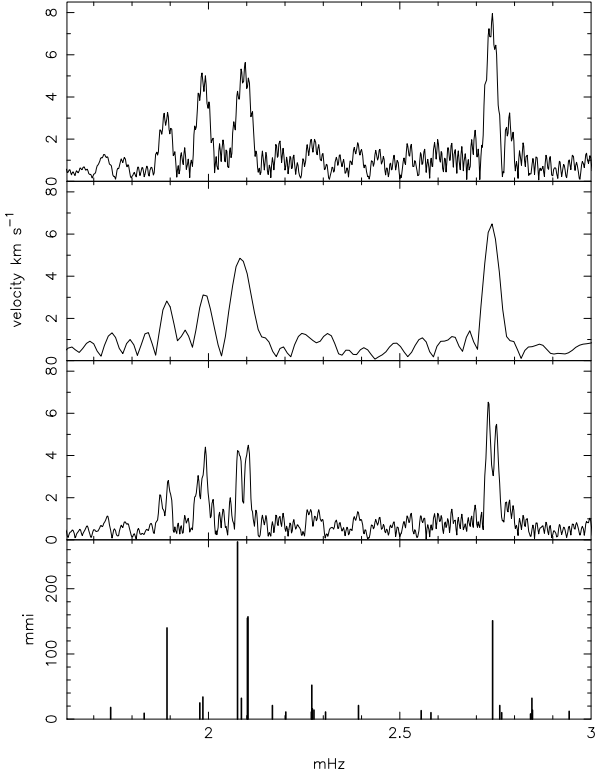}
\caption{Periodograms for WHT, AAT, and complete data sets are shown in the 
top, second, and third panels, respectively.  The frequencies and relative
strengths of peaks found in the photometry periodogram by Kilkenny et al.
\shortcite{k99} are shown in the bottom panel.}
\label{figc}
\end{figure}

\begin{figure}
\epsfig{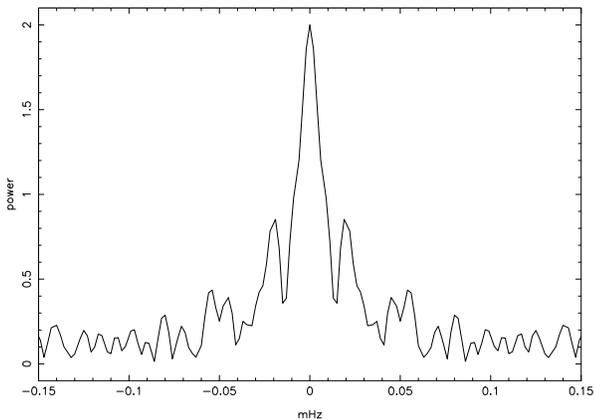}
\caption{Window function for combined AAT and WHT data set.}
\label{figd}
\end{figure}

A search for periods in the velocity variations was performed with a discrete
Fourier transform analysis.  Fig.~\ref{figc} shows the periodograms for the
separate velocity data sets and our complete data set, and the
major photometric frequencies found by Kilkenny et al.
\shortcite{k99} (from their Table~2).
The window function for our data is shown in Fig.~\ref{figd}.

The four major peaks present in the AAT and WHT data become double peaks when
the data is combined.  This raises the question of whether the peaks are
real or the result of aliasing and if the doubling is due to aliasing, which
peaks are the real ones?  The separation of the peaks in the pairs is between
0.019 and 0.028~mHz. The side lobes in the window function are 0.021~mHz
(corresponding to about 13.2 hours) from the central peak.  Aliasing is thus
a possibility.  However, the peak patterns do not resemble that of the window
function. 
The power spectrum and window function of the data set without the WHT
data from the first night, the small set near ${\rm HJD} - 2451676 =0.4$, are
virtually identical to those of the full data set, with slightly broader but
equally separated peaks.  When doing pre-whitening as discussed in the following
paragraph we tried removing peaks in different orders or with the
photometric frequencies and no method removed the second peaks.  Further,
neither peak near 2.7~mHz is at the frequency found using photometry.
Trials using single and multiple sine waves sampled at the observation times,
with and without noise, reproduced the window function at the input
frequency(ies) as expected.
We admit the possibility that the 2.731 and 2.753~mHz peaks are due to aliasing,
but will proceed with the analysis assuming that the splitting is real.

\begin{figure}
\epsfig{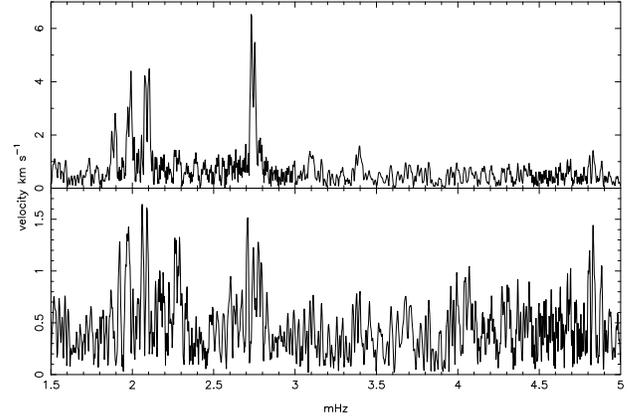}
\caption{Periodogram before (top) and after (bottom) pre-whitening by removing
the six strongest frequency peaks. Note the difference in vertical scale between
panels.}
\label{fige}
\end{figure}

To help determine the relative strengths of the peaks in the periodogram and to
detect peaks which are present only because of aliasing, we ``pre-whitened'' the
data by subtracting successive sine waves from the velocity curve to remove
individual peaks.  The phase for each sine wave was found by cross-correlation
with the velocity curve.  The velocity amplitudes for the sine waves which 
remove the periodogram peaks most cleanly were found by simple trial and error.
The original periodogram and the periodogram with the six strongest peaks
removed are shown in Fig.~\ref{fige}.
The frequencies and velocity amplitudes of the removed peaks are listed in
Table~1. No correction has been made for projection effects.
 The frequency uncertainty was derived using
$$ \delta \omega = \frac{3\pi \sigma _{_N}}{2(N_0)^{1/2}TA} $$
where $A$ is the amplitude, $\sigma_N^2$ is the variance of the noise after
the signal has been removed, $T$ is the length of the data set, and $N_0$ is the
number of velocity data points \cite{ko81,hb86}.
The periodogram of pre-whitened velocity data in Fig.~\ref{fige}
looks similar to the pre-whitened photometric data in Fig.~3 of
Kilkenny et al. \shortcite{k99}, with additional peaks obvious between 1.8 and
3 mHz.  Peaks seen in the photometric data around 4.1 and 4.8 mHz are probably
also present in the velocity data.

\begin{table}
 \begin{minipage}{70mm}
 \caption{Frequencies, periods, and velocities of periodogram peaks as found
while pre-whitening (in order removed). }
  \begin{tabular}{llll}  
\hline
$f$ & $\Delta f$ & $P$ & $v$  \\
mHz & mHz & sec & km s$^{-1}$  \\
\hline
2.731 & 0.001 & 366.2 & 6.1  \\
2.753 & 0.003 & 363.2 & 3.0  \\
2.104 & 0.002 & 475.3 & 4.0  \\
2.076 & 0.002 & 481.7 & 3.9  \\
1.992 & 0.002 & 502.0 & 3.9  \\
1.897 & 0.003 & 527.1 & 2.7  \\
\hline
 \end{tabular}
 \end{minipage}
\end{table}

\begin{figure}
\epsfig{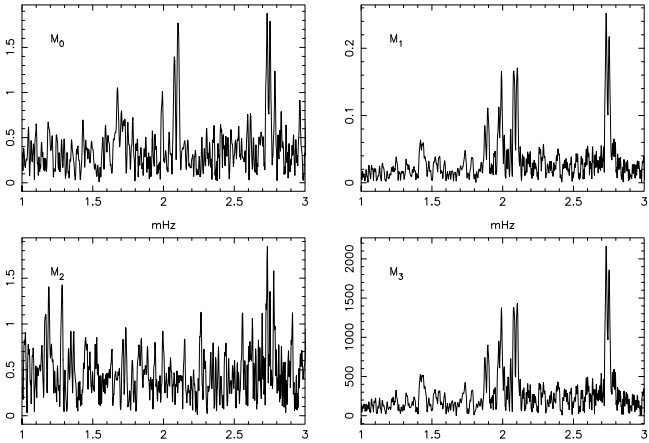}
\caption{Moments of the cross correlation function.}
\label{figf}
\end{figure}

One useful method of mode identification for pulsating stars involves measuring
the moments of the line profiles \cite{b86,a96}. While we are not yet
ready to perform such a test, PG~1605+072, with the largest pulsation amplitudes
known for any sdBV, is an ideal subject to test whether the moments can be
determined for any of the class.  We used the cross correlation function
to approximate the average line profile. For our spectra this is dominated
by the strong H$\gamma$ and H$\delta$ Balmer lines.  An {\sc idl} routine
was used to find the moments $M_0$, $M_1$, $M_2$, and $M_3$, and find their
Fourier transforms.  The transforms are shown in Fig.~\ref{figf}.
The 0th moment corresponds to the
equivalent width of the line.  The first is the line centroid.  The second is
the skewness or lack of symmetry.  The third is the kurtosis, a measure of
how sharply peaked the line is.

The shape of the Fourier transform of the centroid, $M_1$, 
is virtually identical to that of the radial velocity. However, the peak
amplitudes for the centroid are much smaller than those of the the velocities.
We thought that this may happen because the hydrogen line wings are not affected
as strongly by the pulsations as are weak lines or hydrogen line cores. (Our
velocities come from fits to the central peak of the cross correlation function
while the centroid measures the centre of the entire ``line.'')  To test this,
we created synthetic spectra for PG~1605+072 forced to pulsate at various 
frequencies and modes using the programs {\sc bruce} and {\sc kylie}
\cite{t97a,t97b}.  When the spectra included the H $\gamma$ and $\delta$ lines
we found the peak amplitudes for $M_1$ similarly reduced in comparison with
those of the velocities. When the H line regions
of the synthetic spectra were eliminated and the analysis was repeated, the
$M_1$ and velocity amplitudes were comparable, indicating that
the relatively immobile H line wings were the cause of the much smaller
amplitudes found using $M_1$.

The close match between the shapes of the $M_3$ and $M_1$ Fourier transforms
indicates that the ``peakiness'' of the cross correlation function, which
is dominated by the hydrogen line shape, varies with periods and relative
amplitudes identical to the velocity variations.

The $M_0$ and $M_2$ transforms both show peaks at some of the frequencies
present in the $M_1$ and $M_3$ transforms, but with different amplitude
ratios.  This shows that the line equivalent widths and asymmetries vary
differently than the velocity or line kurtoses.  

\section{Discussion}

It is difficult to compare our velocity amplitudes with the  photometric
amplitudes found by other authors,
as there appears to be some shifting of pulsation frequencies.  For example,
the strong peaks in our velocity periodogram at 2.731 and 2.753~mHz do not
correspond well with the peaks in that frequency region in the photometric
periodogram of Kilkenny et~al. \shortcite{k99}, where there is one strong peak
at 2.743 mHz and several much weaker peaks, none of which are at the velocity
peak frequencies.
But still, in both the velocity and photometry periodograms the
major peaks appear in the same general frequency regions: around 1.9, 2.0,
2.1, 2.7, and perhaps 2.3~mHz. 

In addition, Kilkenny et~al. \shortcite{k99} say that some frequencies may
have variable amplitudes (and that some of the small-amplitude peaks in the 
power spectrum may be artifacts due to this variability). This amplitude
variability may be apparent when our velocity data and the Kilkenny et~al.
data are compared.  The strongest peak in the
velocity data is near 2.7 mHz while the strongest in the photometry data is
near 2.1 mHz.  The velocity data from O'Toole et~al. \shortcite{o00} appear to
show amplitude ratios in their periodogram similar to those found in the
Kilkenny et~al. photometric data, though their relatively slow repetition rate
($>$~1 minute per exposure) and the complicated window function for the
O'Toole et~al. data make the comparison uncertain.

Our observations of PG~1605$+$072 took place about three years after the
multi-site photometric campaign to observe it by Kilkenny et~al.
\shortcite{k99} and about 2 years after the O'Toole et~al. \shortcite{o00}
spectroscopic campaign.  The peak amplitude ratio differences we find are
probably due to shifts of power between the major peaks.
Pulsation modes with different $\ell$ and $m$ numbers can produce peaks with
different amplitude ratios for velocity and photometric variations, but
are unlikely to produce the large differences we find.
A campaign of simultaneous spectroscopic and photometric observations would
eliminate the possibility of power shifts between modes over the time period
between obtaining the two types of data.  In addition, the longer observing
runs and wider physical distribution available for the smaller telescopes used
for photometry would help reduce the aliasing problem present with short,
spectroscopy-only programs.

Such a simultaneous campaign would also provide a better chance at
successful mode identification. With modes identified, estimates
of stellar radius and 
luminosity can be made using Baade-Wesselink and related techniques
discussed by Stamford \&\ Watson \shortcite{sw81}.

Mode identification may also be aided by continuing long-term observations
of PG~1605+072.  The rate of period change for the observed pulsations is 
slow enough, ${\rm d}P/{\rm d}t \sim 1.3 \times 10^{-12} {\rm s s^{-1}}$
\cite{k99}, that the shifts in period we observe for the 2.7~mHz peaks
are too large to be the result of the frequency of a single mode changing.
If the shifts are not an artefact of aliasing, they are probably the
result of shifting of power between modes very closely spaced in period,
perhaps different components of a multiplet.  The splitting is about the right
size for this: the frequency difference between modes with successive $m$
values is $\delta \nu \approx \nu_{\rm rot} \frac {1}{\ell(\ell+1)}$, where 
$\nu_{\rm rot}$ is the star's rotational frequency.  If we
guess that the 2.7~mHz peaks are due to $\ell =1$, $m=0,\pm 1$, where the two
peaks we find with the velocity data being $m=\pm 1$ and the peak found
with 1997 and earlier photometric data being $m=0$, then
$\delta \nu = 0.011 \pm 0.002$~mHz, which corresponds to a stellar rotational
period of $12.6 \pm 2.8$~hours,
not far outside the $\Pi < 8.7$~h determined by Heber
et~al.  \shortcite{h99}. (We are not arguing that this {\bf is} an $\ell =1$,
$m=0,\pm 1$ triplet, but rather just that the splitting is about the size
expected for rotational splitting in general for PG~1605$+$072.) If over time
the type of multiplet can be identified (i.e. triplet, quintuplet, etc.) this
would greatly aid in identifying the pulsation modes.
Identifying the pulsation modes would allow analysis of the star using
asteroseismology, providing more precise and accurate estimates of its
physical properties.

\section{Conclusions}
Our two-site campaign of high speed spectroscopic observations of the variable
subdwarf B star PG~1605$+$072 with 4-m class telescopes provides a new
measurement of its radial velocity variations due to pulsation.
We find pulsations frequencies similar
but not identical to those found in earlier photometric studies.  The
relative strengths of the frequency peaks in the power spectrum for our
data are different from those in the photometry data and possibly from those
in an earlier spectroscopic study. The differences in peak strengths and
the small differences in detected frequencies
probably reflect shifts of power between pulsation modes.

A multi-site campaign of simultaneous photometric and spectroscopic observations
combined with continued long-term observations should provide the best chance
to identify the star's pulsation modes.  With the modes identified,
asteroseismology can be used to accurately determine the star's physical
properties, which will be useful in modeling its evolutionary history.

\section*{Acknowledgments}

We are happy to acknowledge financial support from the UK PPARC (grant
Refs PPA/G/S/1998/00019 and PPA/G/O/1999/00058). 

We thank Rich Townsend for providing his {\sc bruce} and {\sc kylie} pulsation
codes.
Thanks to Chris Benn and Raylee Stathakis and others at WHT and AAT for tracking
down information on the timing calibrations of the observations.

\end{document}